\documentclass{aastex631}

\usepackage{xcolor}

\usepackage{amsmath}
\usepackage{booktabs}
\usepackage{multirow}

\begin{document}

\title{Characterizing the emission region property of blazars}

\correspondingauthor{Junhui Fan, Hubing Xiao}
\email{fjh@gzhu.edu.cn, hubing.xiao@shnu.edu.cn}

\author[0000-0002-5929-0968]{Junhui Fan}
\affiliation{Center for Astrophysics, Guangzhou University, Guangzhou 510006, China}
\affiliation{Key Laboratory for Astronomical Observation and Technology of Guangzhou, Guangzhou, 510006, China}
\affiliation{Astronomy Science and Technology Research Laboratory of Department of Education of Guangdong Province, \\
Guangzhou, 510006, China}

\author[0000-0001-8244-1229]{Hubing Xiao}
\affiliation{Shanghai Key Lab for Astrophysics, Shanghai Normal University, Shanghai, 200234, China}

\author{Wenxin Yang}
\affiliation{Center for Astrophysics, Guangzhou University, Guangzhou 510006, China}
\affiliation{Key Laboratory for Astronomical Observation and Technology of Guangzhou, Guangzhou, 510006, China}
\affiliation{Astronomy Science and Technology Research Laboratory of Department of Education of Guangdong Province, \\
Guangzhou, 510006, China}

\author{Lixia Zhang}
\affiliation{Center for Astrophysics, Guangzhou University, Guangzhou 510006, China}
\affiliation{Key Laboratory for Astronomical Observation and Technology of Guangzhou, Guangzhou, 510006, China}
\affiliation{Astronomy Science and Technology Research Laboratory of Department of Education of Guangdong Province, \\
Guangzhou, 510006, China}

\author{Anton A. Strigachev}
\affiliation{Institute of Astronomy and National Astronomical Observatory, Bulgarian Academy of Sciences,\\
 72 Tsarigradsko shosse Boulevard, 1784 Sofia, Bulgaria}

\author[0000-0002-0766-864X]{Rumen S. Bachev}
\affiliation{Institute of Astronomy and National Astronomical Observatory, Bulgarian Academy of Sciences,\\
 72 Tsarigradsko shosse Boulevard, 1784 Sofia, Bulgaria}

\author{Jianghe Yang}
\affiliation{Department of Physics and Electronics Science, Hunan University
of Arts and Science, Changde 415000, China}
\affiliation{Center for Astrophysics, Guangzhou University, Guangzhou 510006, China}



\begin{abstract}
The studies and constraints on the emission region are crucial to the blazar radiation mechanism.
Yet the previous works mainly focus on individual sources.
In this work, we make use of the largest and the latest spectral energy distribution (SED) fitting results in the literature to statistically study the blazar emission region property in the framework of leptonic one-zone.
Our results reveal
(1) FSRQs show lower electron energy ($\gamma_{\rm p} \lesssim 1.6 \times 10^{3}$) than BL Lacs and tend to have a stronger magnetic field ($B$) and smaller electron-to-magnetic energy ratio ($U_{\rm e}/U_{\rm B}$) than BL Lacs;
(2) we find the electro-magnetic equipartition would rather happen in the jets of BL Lacs than happen in the jets of FSRQs;
(3) there are 682 blazars with a magnetic field weaker critical value of generating the Kelvin-Helmholtz instability, thus one-third of the blazars in our sample are able to produce this instability;
(4) the distance ($d_{\rm em}$) between the emission region and the central black hole (BH) is in the scale of $\sim$0.1 pc, the location of the emission region may be evenly distributed inside and outside the broad line region (BLR).

\end{abstract}


\section{Introduction} \label{sec:introduction}
Active galactic nuclei (AGNs), one of the most popular extragalactic objects in astronomy, emit radiation in the overall electromagnetic spectrum.
The ultimate energy source of AGNs is believed to be the gravitational potential of supermassive black holes (BHs), which are embedded in its center, rather than the nuclear fusion of stars \citep{Lynden1969Nat}.
An accretion disk can be formed surrounding the BHs by matter losing angular momentum before falling onto the BHs \citep{Rees1984ARAA, Cao2013ApJ}.
AGNs are divided into radio-loud and radio-quiet ones according to their relative radio emission intensity compared to their optical emission intensity \citep{Strittmatter1980, Kellermann1989, Xiao2022PASJ}.
This dichotomy is mainly caused by the presence of a strong relativistic and collimated jet in the radio-loud AGNs \citep{Urry1995}.
The jet is so powerful that dominates the entire emission of radio-loud AGNs, but the mechanism of jet launching is still controversial.
\citet{Blandford1977} suggested that the jet is powered by extracting the rotational energy of BH (B-Z process), while \citet{Blandford1982} suggests that the jet is powered by the rotational energy of accretion disk (B-P process) \citep[see also in][]{Xiong2014, Xiao2022ApJ_1, Zhang2022ApJ}.

Blazars, as an extreme subclass of radio-loud AGNs, exhibit distinctive observational properties.
The rapid and large amplitude variability, high and variable polarization, strong and variable $\gamma$-ray emissions, and apparent superluminal motion of blazars have been observed and studied \citep{Wills1992, Urry1995, Villata2006, Fan2002, Fan2014, Fan2021, Gupta2016, Xiao2019, Xiao2020, Xiao2022MNRAS, Abdollahi2020}.
These properties are believed as consequences of a Doppler beaming effect due to a small viewing angle ($\theta$) between the jet axis and line of sight \citep[eg.,][]{Ghisellini1993, Fan2013, Pei2016, Xiao2020}.
The Doppler beaming effect is usually pronounced as the time dilation and intensity amplification through a Doppler factor ($\delta=[\Gamma(1-\beta {\rm cos}\theta)]^{-1}$, where $\Gamma$ is the bulk Lorentz factor and $\beta$ is the jet speed in units of the speed of light, $c$), which can be estimated indirectly \citep{Ghisellini1993, Readhead1994, Fan2005ChJAS, Fan2013, Liodakis2018}.
There are two subclasses of blazars, namely flat spectrum radio quasars (FSRQ) and BL Lacertae objects (BL Lacs).
The former one characterizes an optical spectrum with strong emission lines (rest-frame equivalent width, ${\rm EW > 5 \AA}$), while the latter one demonstrates no or weak emission features (${\rm EW < 5 \AA}$) \citep{Urry1995, Scarpa1997}.
Broadband studies illustrate a typical two-hump spectral energy distribution (SED), the lower energy bump ranges from radio to X-ray band and peaks at the infrared (IR) to X-ray band, which is believed to be the synchrotron emission of the relativistic electrons in the jet, the higher energy bump starts from the X-ray band to the $\gamma$-ray band and peaks at the X-ray to GeV $\gamma$-ray band \citep{Abdo2010, Fan2016, Paliya2021, Yang2022ApJS, Yang2023SCPMA}.
The radiation mechanism of the higher energy bump is still controversial, the leptonic model suggests that the higher energy bump is attributed to the inverse Compton (IC) process \citep{Blandford1979, Sikora1994, Sokolov2005, Abramowski2015, Xue2019ApJ_1, Tan2020, Wang2022PASP}, while the hadronic model interprets it through the proton synchrotron radiation and secondary particle cascade \citep{Mucke2001, Dimitrakoudis2012, Diltz2015, Cerruti2015MNRAS, IceCube2018, Xue2021, Wang2022PRD}.  

Broadband SED modelling is an efficient approach to investigate the jet properties, eg., constraining the magnetic field ($B$) of the emission region, the Doppler factor, the emission region size ($R$), and the electron energy distribution (EED) \citep{Massaro2004a, Tramacere2011, Ghisellini2014, Ghisellini2015MNRAS, Chen2017ApJ}.
However, there are degeneracies between these critical parameters \citep{Kubo1998ApJ, Ghisellini2014}.
Additional methods are employed to further constrain these parameters, for instance, the variability time scale is used to give an upper limit of $R\leq c \Delta t \delta/(1+z)$ due to the causality; 
the Owens Valley Radio Observatory (OVRO) made efforts of estimating $\delta$ through radio lightcurve of blazar flare/outburst \citep{Liodakis2018}; 
the Monitoring of Jets in AGN with VLBA Experiments (MOJAVE) program monitors the radio brightness, polarization variation and apparent motion \citep{Lister2015, Lister2019}, and the apparent speed is an indicator of $\delta$ \citep{Zhang2008, Xiao2019, Xiao2020};
The polarization observation is critical to the study of blazar magnetic field.
The observed polarization shows a frequency-dependence of rotation measure (RM) in 3C 273 and suggests the magnetic field may be structured helically on a large scale \citep{Wardle2018Galax, Hovatta2019AA}.
Recently, constraining magnetic field strength with simultaneously observed optical and radio linear polarization and circular polarization was suggested by \citet{Liodakis2022MNRAS}, who proposed a formula that the magnetic field strength is proportional to the square of circular polarization and inversely proportional to the linear polarization.
Based on this, they rejected that the high-energy emission models requiring high magnetic field strength and a low positron fraction.

It is also accessible to determine magnetic field, Doppler factor, emission region size, etc, via SED features. 
\citet{Kubo1998ApJ} considered a population of relativistic electrons, which forms an EED of broken power-law with a breakpoint at $\gamma_{\rm p}$, and assumed that the lower-energy component SED peaks at a frequency corresponding to that radiated by the electrons with $\gamma_{\rm p}$.
Coupling the assumption with the synchrotron and IC radiation mechanism, they calculated the strength of the magnetic field and the electron Lorentz factor with the energy and intensity of the two SED peaks.  
Similarly, \citet{Chen2018} applied the same method with a log-parabola EED and obtained formulas to calculate these parameters describing jet properties.
Benefiting from the observations of \textit{Fermi} Large Area Telescope (LAT), a large number of blazars have been discovered \citep{Abdollahi2020}, and the study of blazar comes to its era of prosperity.

In this work, taking the advantage of the latest and largest blazar SED fitting results of \citet{Yang2022ApJS} and \citet{Yang2023SCPMA}, we aim to investigate the property of blazar emission region and put constraints on those critical parameters.
This paper is organized as follows:
In section \ref{sec:sample} we present our sample;
Our method, analysis and results will be presented in Section \ref{sec:result};
The discussions in Section \ref{sec:discussion};
Our conclusions will be given in Section \ref{sec:conclusion}.

\section{Sample} \label{sec:sample}
To study the blazar emission region and constrain its relevant parameters, we need to connect these parameters with observational quantities.
We collected a sample of \textit{Fermi} blazars with available synchrotron peak frequency ($\log \nu_{\rm sy}$) and the corresponding luminosity ($\log L_{\rm sy}$) from \citet{Yang2022ApJS} and the IC peak frequency ($\log \nu_{\rm IC}$) and luminosity ($\log L_{\rm IC}$) from \citet{Yang2023SCPMA}.
In total, we have 2708 sources, including 759 FSRQs, 1141 BL Lacs, and 808 blazar candidates of uncertain type (BCUs).
There are 1791 sources in our sample with available redshift, including 750 FSRQs with an average value of $\langle z_{\rm F} \rangle= 1.201 \pm 0.647$, 843 BL Lacs with average value $\langle z_{\rm B} \rangle = 0.528 \pm 0.501$, and 198 BCUs with average value $\langle z_{\rm U} \rangle= 0.919 \pm 0.738$.
In this sample, there are 512 blazars are associated with sources in \citet{Liodakis2018} and thus with available Doppler factor ($\delta$), with an average value of $\langle \delta_{\rm F} \rangle = 17.39 \pm 13.42$ for FSRQs, an average value of $\langle \delta_{\rm B} \rangle = 11.26 \pm 10.29$ for BL Lacs, and an average value of $\langle \delta_{\rm U} \rangle = 12.50 \pm 13.37$ for BCUs.

\begin{table}[htbp]
\centering
\caption{The magnetic field strength and electron energy for 2708 \textit{Fermi} blazars.}
\label{para}
\begin{tabular}{lccccccccccc}
\toprule
4FGL Name & Class\_1 & Class\_2 & z & $\delta$ & $\log \nu_{\rm sy}$ & $\log \nu_{\rm IC}$ & P1 & $\log B$ & $\log \gamma_{\rm p}$  & $\log U_{\rm e}$ & $B_{\rm c}$\\
(1) & (2) & (3) & (4) & (5) & (6) & (7) & (8) & (9) & (10) & (11) & (12) \\
\hline
\midrule 
J0001.2-0747 &	BLL	& ISP &   &	  &	14.1 &	23.14	& -0.12	&      -2.25        &	4.46  &	1.82  &	28.82 \\
J0001.2+4741 &	BCU	& ISP &   &	  &	14.1 &	22.50	& -0.09	& $-1.56 \sim 1.64$ & $4.14 \sim 2.54$ & $0.68 \sim -4.26$ & \\	
J0001.5+2113 &	FSRQ &	LSP	& 1.11	&	& 13.2 & 20.62 &	-0.18 &	2.77 &	1.47 &	-5.86	& 0.004 \\
J0002.4-5156 & BCU	 &  HSP &   &    &	15.7 & 23.99 & -0.09 & $0.15 \sim 0.25$ & $4.08 \sim 4.03$ & $-2.29 \sim -2.44$ &	\\
J0003.1-5248 & BCU	& HSP & & &	15.9 & 24.31 & -0.07 & $0.23 \sim 0.13$ & $4.14 \sim 4.19$ & $-2.24 \sim -2.09$ &	\\
...             & ...   & ...   & ...   &...   & ...    & ...    & ...   & ...                  & ...               &   ...    &  ...  \\
\hline
\bottomrule
\end{tabular}
\tablecomments{
column (1) gives the sources name;
column (2) is the spectral classification;
column (3) is the classification based on synchrotron peak frequency \citep{Fan2016, Yang2022ApJS};
column (4) gives the redshift;
column (5) is the Doppler factor from \citet{Liodakis2018};
column (6) and column (7) gives the synchrotron and IC peak frequencies from \citet{Yang2022ApJS, Yang2023SCPMA};
column (8) is the spectral curvature from \citet{Yang2022ApJS};
column (9) is the magnetic field in units of Gs given in this work;
column (10) is the energy of electrons contributing most to the synchrotron peak;
column (11) is the energy density of the electrons;
column (12) is the critical magnetic field strength.
There are only 5 items displayed here, the table is available in its entirety in machine-readable form.
}
\end{table}

\section{Method and Results} \label{sec:result}
\subsection{The basics of one-zone leptonic model}
In the leptonic frame, the higher energy emission results from the inverse Compton scattering of internal or external soft photons, namely the synchrotron self-Compton (SSC) and external Compton (EC) process.
The SSC process is believed to dominate the radiation in BL Lacs jets, while the EC process dominates over the SSC process for FSRQs jets.
The synchrotron power is mainly produced by those electrons with Lorentz factor of $\gamma_{\rm p}$ and contribute most to the synchrotron peak, and the synchrotron peak frequency in the observer frame is given by
\begin{equation}
    \nu_{\rm sy} = 3.7 \times 10^{6}\, \gamma_{\rm p}^{2} \, B\, \frac{\delta}{1+z} \, {\rm Hz},
\label{nu_sy}
\end{equation}
where $B$ in units of Gs \citep{Tavecchio1998}.
Correspondingly, the synchrotron peak luminosity in the observer frame is expressed as \citep{Chen2018}
\begin{equation}
    L_{\rm sy} = 4\pi \, \frac{16}{9} \pi R^{3}\, \frac{\sigma_{\rm T} c}{8\pi}\, U_{\rm B}\, N_{\rm 0}\, \gamma_{\rm p}^{3}\, \delta^{4},
\label{L_sy}
\end{equation}
where $\sigma_{\rm T}$ is the Thomson cross section, $U_{\rm B}$ is the magnetic field energy density ($U_{\rm B} = B^{2}/8\pi$), $N_{\rm 0}$ is the normalization parameter of the EED.
Following \citet{Chen2018}, a three-parameter log-parabolic function is employed to describe the EED in this work
\begin{equation}
    N(\gamma) = N_{\rm 0} \left( \frac{\gamma}{\gamma_{\rm p}} \right)^{-3} 10^{-2b\log(\gamma/\gamma_{\rm p}) },
\label{n}
\end{equation}
$b$ is the curvature and has a relation with the synchrotron bump, $b\simeq 5 P_{\rm 1}$ \citep[see, eg.,][]{Massaro2006, Chen2014}.
This EED is only phenomenologically assumed to follow the log-parabola shape of SED, without taking into account the evolution due to injection and cooling effects. 

Moreover, if the electrons in the Thomson regime, the peak frequency of the SSC component is 
\begin{equation}
    \nu_{\rm SSC} = \frac{4}{3}\, \gamma_{\rm p}^{2} \, \nu_{\rm sy}.
\label{nu_ssc}
\end{equation}
In the case of the EC process, soft photons are fed externally, the peak frequency is given by
\begin{equation}
    \nu_{\rm EC} = \frac{4}{3}\, \gamma_{\rm p}^{2} \, \nu_{\rm ext} \, \frac{\Gamma \delta}{1+z},
\label{nu_ec}
\end{equation}
where, $\nu_{\rm ext}$ is the frequency of external photons, $\nu_{\rm ext} = 2.46 \times 10^{15} \, {\rm Hz}$ for the case of external photons coming from the broad line region (BLR) and $\nu_{\rm ext} = 7.7 \times 10^{13} \, {\rm Hz}$ for the case of external photons coming from the dusty torus (DT) \citep{Tavecchio2008MNRAS386, Ghisellini2015MNRAS}.

\subsection{The magnetic field and the Lorentz factor of electrons}
\begin{figure}
\centering
\includegraphics[scale=0.85]{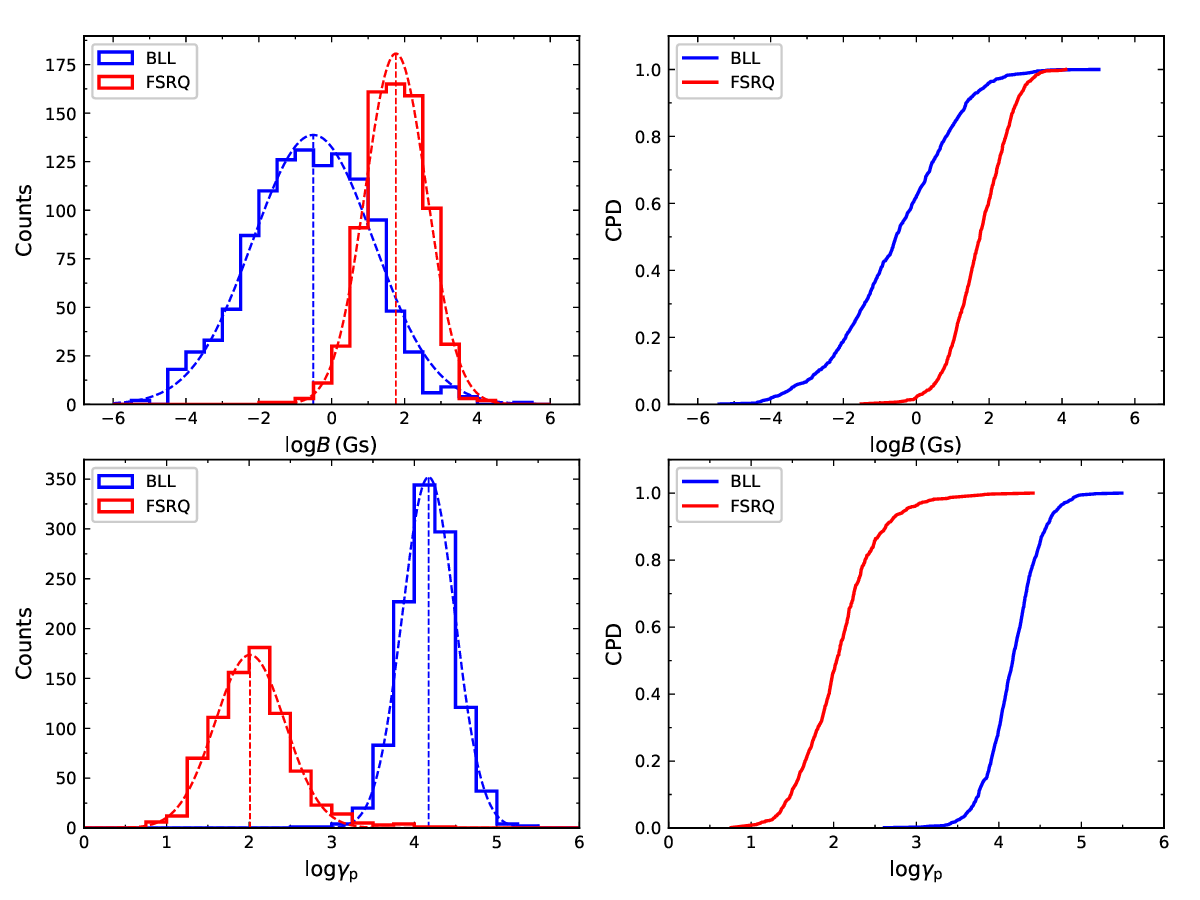}
\caption{The distributions of $B$ and $\gamma_{\rm p}$. 
The left panel gives the distributions of the two parameters, the dashed curve stands for the Gaussian fit of the histogram.
The right panel gives the corresponding cumulative probability distributions (CPD). 
The red colour stands for FSRQs and the blue colour stands for the BL Lacs throughout this paper}.
\label{dis_bg}
\end{figure}

The magnetic field and the electron energy are crucial to the radiation model of blazar jets.
In a leptonic model, one can estimate these two parameters, $B$ and $\gamma_{\rm p}$, through equations \ref{nu_sy}, \ref{nu_ssc} and \ref{nu_ec} in both SSC and EC processes.
We calculated $B$ and $\gamma_{\rm p}$ for the 1900 blazars, including 1141 BL Lacs and 759 FSRQs, the SSC process is assumed to be the emission case of BL Lacs and the EC process is considered for the FSRQs.
An average replacement is used for those sources without available redshift or Doppler factors, $\langle z_{\rm F} \rangle = 1.21$ and $\langle \delta_{\rm F} \rangle =17.47$ for FSRQs, $\langle z_{\rm B} \rangle = 0.528$ and $\langle \delta_{\rm B} \rangle =11.26$ for BL Lacs.
In addition, for the EC process in FSRQs, we consider the external photons from the DT for those HSPs, while the external photons with an origination of the BLR for those LSPs and ISP.
In this case, we have 754 FSRQs using the model of external soft photons coming from the BLR, this is consistent with the assumption of soft photon origin for FSRQs in literature \citep[e.g.,][]{Tan2020}, and also consistent with the fact that the FSRQs show significant broad emission lines and the emission lines contribute to the EC component significantly \citep{Xiao2022ApJ_2}.
The rest of the 5 FSRQs are considered as HSPs, and the HSPs are naturally considered as TeV candidates \citep{Zhu2023ApJS}. 
The TeV emission could be severely absorbed by interacting with BLR soft photons, thus we assume these 5 FSRQs with soft photons from the DT.
The results of $\log B$ and $\log \gamma_{\rm p}$ are listed in columns (9) and (10) of Table \ref{para}, and are displayed in Figure \ref{dis_bg}. 
The Gaussian fit is applied to the distributions, and the fitting results give 
a mean value $\log B^{\rm B}=-0.51$ with a standard deviation of 1.66 for BL Lacs 
and a mean value $\log B^{\rm F}=1.76$ with a standard deviation of 0.85 for FSRQs;
a mean value $\log \gamma_{\rm p}^{\rm B}=4.17$ with a standard deviation of 0.32 for BL Lacs 
and a mean value $\log \gamma_{\rm p}^{\rm F}=2.01$ with a standard deviation of 0.43 for FSRQs.
A two-sample Kolmogorov–Smirnov (K–S) test, is employed to test whether each parameter for these two subclasses is from the same parent distribution.
In the K-S test, the probability smaller than the critical value ($p=0.05$) would be used to reject the null hypothesis, which is that the two distributions are coming from the same parent distribution.
Our results of K-S tests show $p \sim 0$ for both $\log B$ and $\log \gamma_{\rm p}$ distributions, suggest the BL Lac $\log B^{\rm B}$ distribution and the FSRQ $\log B^{\rm F}$ distribution come from different parent distributions, as well as for the BL Lac $\log \gamma_{\rm p}^{\rm B}$ distribution and the FSRQ $\log \gamma_{\rm p}^{\rm F}$ distribution.

Besides, we calculate the range of both $\log B$ and $\log \gamma_{\rm p}$ for the 808 BCUs in our sample, the limits are obtained by assuming the seed photons come from either the SSC or from the EC.
During the calculation, the average replacement is used for those BCUs without available redshift and Doppler factors.
The results are also listed in columns (9) and (10) of Table \ref{para}.

\subsection{The emission region of blazars}
\begin{figure}
\centering
\includegraphics[scale=0.6]{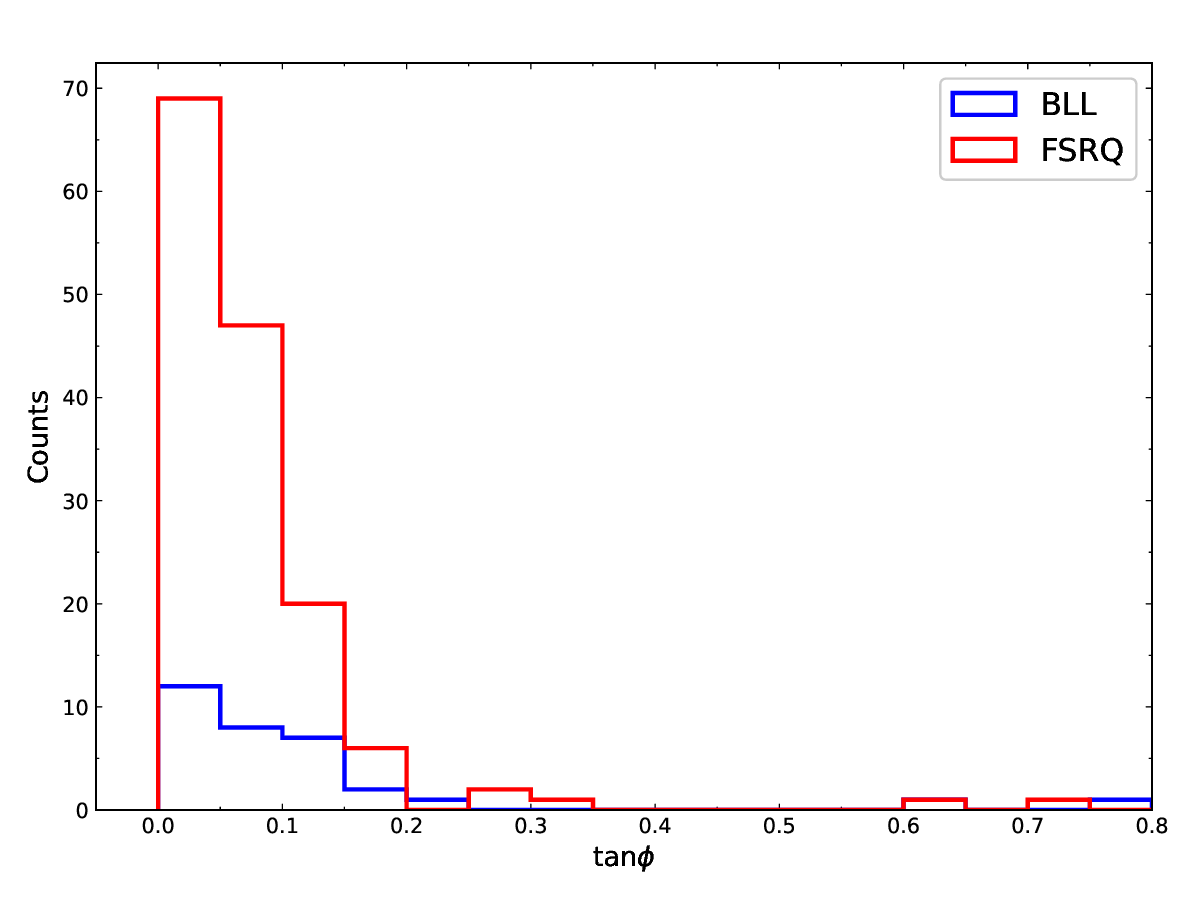}
\caption{The distribution of $\tan \phi$}
\label{dis_phi}
\end{figure}

\begin{figure}
\centering
\includegraphics[scale=0.85]{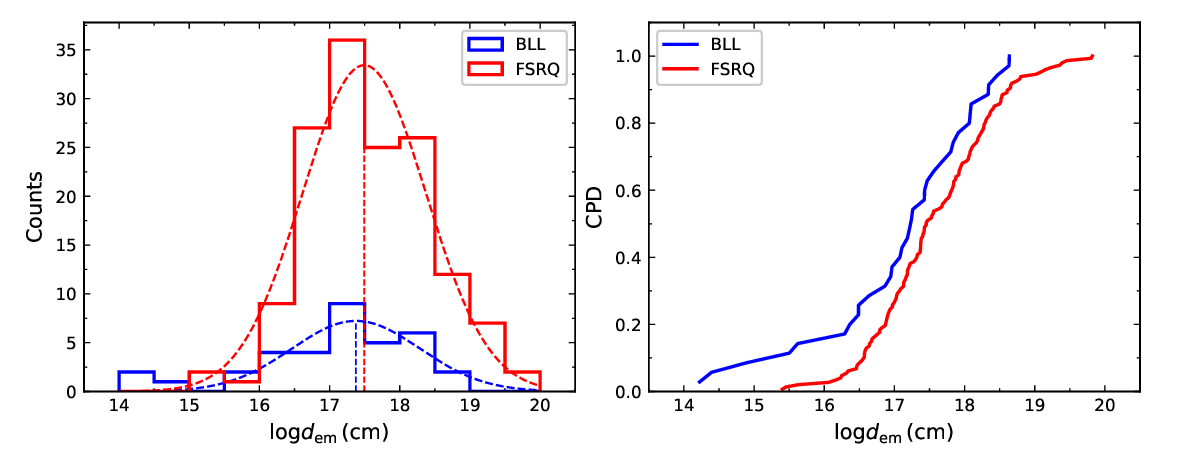}
\caption{The distribution of $\log d_{\rm em}$}
\label{dis_d_em}
\end{figure}

The jet emission region size ($R$) is usually constrained by a causality reason, which is the variability time scale, and expressed as 
\begin{equation}
    R \leq c \, \Delta t \, \frac{\delta}{1+z}.
\label{r}
\end{equation}
$R$ is easily obtained if we assume a variability timescale $\Delta t$ for those blazars with available $z$ and $\delta$.
Throughout this paper, we assume $\Delta t = 1 \, {\rm day}$ \citep{Nalewajko2013MNRAS, Fan2013} to estimate $R$.
Consider a constant and symmetric jet geometry, if the full jet cross section is responsible for the emission region diameter, then the distance from the central SMBH to the emission region $d_{\rm em}$ and the emission region size $R$ are correlated as
\begin{equation}
R = d_{\rm em} \tan\! \phi,
\label{Eq_psi}
\end{equation}
where $\phi$ is the semi-aperture opening angle of the jet \citep{Acharyya2021}.
$\phi$ is an undetectable quantity, the value of $\tan \phi =0.1$ is fixed to study the blazar jet power in \citet{Ghisellini2009} and $\tan \phi \lesssim 0.25$ is suggested in \citet{Dermer2009}.
In this work, we can assume this opening angle $\phi$ to be close to the viewing angle $\theta$ for blazars ($\theta \simeq \phi$) due to the fact that the blazar jet is pointing at observers.
Actually, $\theta$ can be estimated by using the apparent velocity of the resolved jet components and expressed as 
\begin{equation}
    \tan\! \phi \simeq \tan\! \theta = \frac{2 \beta_{\rm app}}{\beta^{2}_{\rm app} + \delta^{2} - 1},
\end{equation}
where $\beta_{\rm app}$ is the apparent velocity of the jet component \citep{Kellermann2004, Liodakis2018, Xiao2019}, which is usually observed via the Very Long Baseline Interferometry (VLBI) technique \citep{Lister2009, Lister2019, Lister2021}.
We manage to collect $\theta$ from \citet{Liodakis2018} for 182 blazars (35 BL Lacs and 147 FSRQs) of our sample and list them in column (5) of Table \ref{distance}, and show them in Figure \ref{dis_phi}.
We notice that there are 142 of 147 FSRQs with $\tan\! \theta < 0.25$, taking 96.9\%, there are 30 of 35 BL Lacs with $\tan\! \theta < 0.25$, taking 85.7\%, .
Then we calculate the $d_{\rm em}$ for these 182 sources and list the results in column (6) of Table \ref{distance}, and show them in Figure \ref{dis_d_em}.
The distribution of the emission region distance gives 
a mean value $\log d_{\rm em}^{\rm B}=17.37$ with a standard deviation of 0.91 for BL Lacs 
and $\log d_{\rm em}^{\rm F}=17.49$ with a standard deviation of 0.88 for FSRQs.
A K-S test result of $0.31$ suggests that the $\log d_{\rm em}^{\rm B}$ and $\log d_{\rm em}^{\rm F}$ could come from the same distribution.

\begin{table}[htbp]
\centering
\caption{The location of the emission region for 182 blazars.}
\label{distance}
\begin{tabular}{lccccccc}
\toprule
4FGL Name & Class\_1 & z & $\delta$ & $\theta$ & $\log d_{\rm em}$ & $\log d_{\rm em}\, {\rm (G15)}$ & $\log L_{\rm BLR}$   \\
(1) & (2) & (3) & (4) & (5) & (6) & (7) & (8)  \\
\hline
\midrule 
J0006.3-0620	& BLL	& 0.346676	& 6.96	  & 8.25	  & 16.97	&        & 43.60	  \\
J0017.5-0514	& FSRQ	& 0.227	    & 12.02	  & 0.72	  & 18.31	& 16.17  & 43.77	  \\
J0019.6+7327	& FSRQ	& 1.781	    & 7.84	  & 7.32	  & 16.75	&        & 45.62	  \\
J0051.1-0648	& FSRQ	& 1.975	    & 5.61	  & 7.27	  & 16.58	&        & 46.11	  \\
J0102.8+5824	& FSRQ	& 0.644	    & 18.51	  & 2.39	  & 17.84	& 16.75  & 45.04	  \\
...             & ...   & ...       & ...     &...        & ...     &        & ...        \\
\hline
\bottomrule
\end{tabular}
\tablecomments{
column (1) gives the sources name;
column (2) is the spectral classification;
column (3) gives the redshift;
column (4) and (5) are the Doppler factor and the viewing angle from \citet{Liodakis2018};
column (6) the estimated $\log d_{\rm em}$ from this work;
column (7) the estimated $\log d_{\rm em}$ in \citet{Ghisellini2015MNRAS};
column (8) gives the luminosity of the BLR from \citet{Xiao2022ApJ_1}.
There are only 5 items displayed here, the table is available in its entirety in machine-readable form.
}
\end{table}

\section{Discussions} \label{sec:discussion}
\subsection{The energy density of the relativistic electron population}
In the system of a blackhole-based jet, the particle-field relations depend on uncertain jet formation, particle acceleration, and radiation mechanisms \citep{Dermer2014ApJ}.
It is natural that systems with interacting components often tend to equipartition.
As a synchrotron source, blazar jet contains relativistic electrons with some energy density $U_{\rm e}$ and a magnetic field whose energy density is $U_{\rm B} = B^{2}/(8\pi)$.
In order to avoid involving those poorly understood microphysics processes in the jets, \citet{Dermer2014ApJ} assumed a condition that the equipartition between magnetic-field and non-thermal electron energy densities holds for blazar jets, as it was used in the analysis of a  wide variety of astrophysical systems; see \citet{Burbidge1959ApJ} for the radio lobes, \citet{Pacholczyk1970ranp} for a study of radio sources, and \citet{Beck2005AN} for a theoretical equipartition study of synchrotron observations.
On large scales, the equipartition also corresponds to the minimum jet power condition as suggested by \citet{Ghisellini2001MNRAS}, in which the minimum jet power is required to produce the radiation we observe at all wavelengths.
The simplest equipartition relation is $U_{\rm e} = \xi U_{\rm B}$, $\xi$ is close to the unity \citep{Ghisellini2001MNRAS, Dermer2014ApJ}.
It is believed that the jet is in global equipartition between $U_{\rm e}$ and $U_{\rm B}$, however, local out-of equipartition is also allowed to explain peculiar observation phenomenon.
The very high energy (VHE) detection of 3C 279 \citep{MAGIC2008Sci} required either emission from leptons far out of equilibrium accompanied by poor fits to the X-ray or synchrotron data in the leptonic frame.
In 2018, \textit{Fermi} observed a characteristic peak-in-peak variability pattern on time scales in minutes of 3C 279 resulting from magnetic reconnection \citep{Shukla2020NatCo}. 

In Figure \ref{dis_bg}, we notice that the FSRQs show a stronger magnetic field than the BL Lacs, while the latter shows a higher energy of electrons than the former.
This is consistent with the typical blazar radiation mechanism paradigm, in which FSRQ contains a photon-rich environment so that the energy of relativistic electrons in the emission blob could efficiently dissipate via the IC process.
And, BL Lacs is believed to be located in a photon-starving environment, thus the accelerated relativistic electrons can be well preserved.
Moreover, the distribution of $\log \gamma_{\rm p}$ shows a clear separation between the FSRQs and the BL Lacs.
This result reveals the energy of the relativistic electrons is distributed in a wide range, which contains large variance for different types of blazars, the FSRQs taking the side of lower energy while the BL Lacs taking the side of higher energy.
We suggest using the intersection point value, $\log \gamma_{\rm p}= 3.20 \pm 0.01 \, (\gamma_{\rm p}\simeq 1.6 \times 10^{3})$, of the two Gaussian profiles to divide FSRQs and BL Lacs.

\begin{figure}
\centering
\includegraphics[scale=0.85]{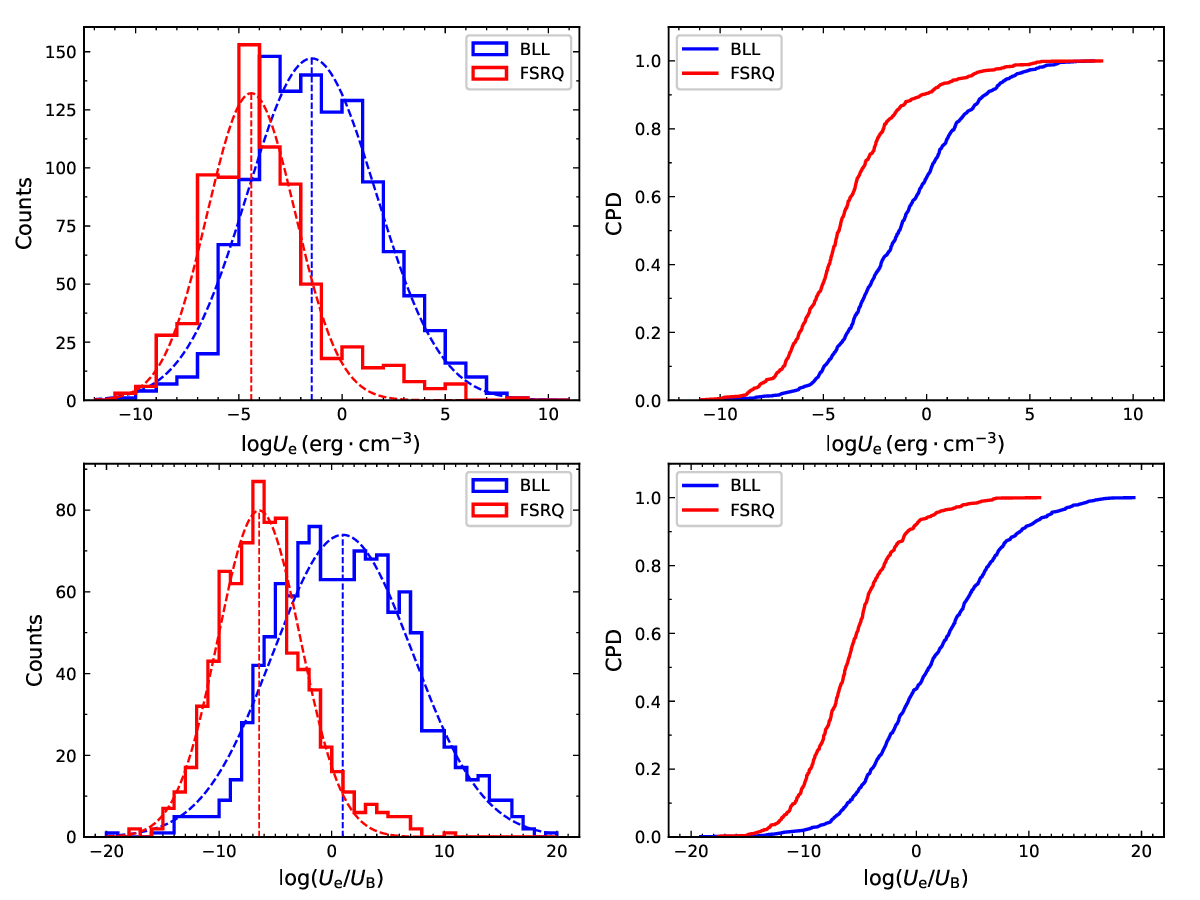}
\caption{The distributions of $\log U_{\rm e}$ and $\log (U_{\rm e}/U_{\rm B})$}
\label{dis_u}
\end{figure}

With known $B$, $\gamma_{\rm p}$, and $N(\gamma)$, one can obtain the magnetic energy density $U_{\rm B}$ and the electron energy density
\begin{equation}
    U_{\rm e} = \int_{\gamma_{\rm min}}^{\gamma_{\rm max}} \gamma \, m_{\rm e}c^{2} N(\gamma) {\rm d}\gamma, 
\label{u_e}
\end{equation}
where the minimum Lorentz $\gamma_{\rm min} = 10$ and the maximum Lorentz factor $\gamma_{\rm max}=1\times 10^{6}$ of electrons is assumed, $m_{\rm e}$ is the rest mass of electron.
Figure \ref{dis_u} shows the distribution of $\log U_{\rm e}$ and the ratio $\log (U_{\rm e}/U_{\rm B})$.
The Gaussian fit gives results of 
a mean value $\log U_{\rm e}^{\rm B}=-1.47$ with a standard deviation of 3.12 for BL Lacs 
and $\log U_{\rm e}^{\rm F}=-4.40$ with a standard deviation of -2.14 for FSRQs;
a mean value $\log (U_{\rm e}^{\rm B}/U_{\rm B}^{\rm B})=0.98$ with a standard deviation of 6.23 for BL Lacs 
and a mean value $\log (U_{\rm e}^{\rm F}/U_{\rm B}^{\rm F})=-6.45$ with standard deviation of 3.68 for FSRQs.
K–S tests of $\log U_{\rm e}$ and of $\log (U_{\rm e}/U_{\rm B})$ for BL Lacs and FSRQs, and give both results of $p\sim0$.
The results suggest $\log U_{\rm e}^{\rm B}$ and $\log U_{\rm e}^{\rm F}$ are from different distributions, as well as for the distributions of $\log (U_{\rm e}^{\rm B}/U_{\rm B}^{\rm B})$ and $\log (U_{\rm e}^{\rm B}/U_{\rm B}^{\rm F})$.
Based on our results shown in Figure \ref{dis_bg} and \ref{dis_u}, we suggest BL Lacs have an averagely larger energy and energy density of electron distribution, meanwhile, it also shows a larger electron-to-magnetic energy ratio than FSRQs.

Based on the distribution of $\log (U_{\rm e}/U_{\rm B})$ in Figure \ref{dis_u}.
It is clear that most of the FSRQs are away from the value $\log (U_{\rm e}/U_{\rm B})=0$, which is the condition of equipartition between magnetic-field and non-thermal electron energy densities, while the BL Lacs have a mean value near $\log (U_{\rm e}/U_{\rm B})=0$.
Thus, our results suggest that the BL Lacs stay in a `quasi-equipartition' state, while the FSRQs do not.

\subsection{The Kelvin–Helmholtz instability}
Variation, one of the characterizing properties of blazars, has been observed across all frequencies and time scales \citep[eg.,][]{Urry1996ASPC, Dermer1999APh, Fan1999MNRAS, Singh2020AN, Webb2021Galax, Otero2022MNRAS, Amaya2022ApJ}.
Blazar variability time scales are observed from years to months, to days, and even to minutes \citep{Wagner1995ARAA, Fan1998ApJ, Fan2018AJ, Aharonian2007ApJ, Albert2007ApJ}.
The basic idea of the mechanism for these different time scales of variabilities is that a disturbance created near the black hole travels outward with a Lorentz factor $\Gamma$ and radiates energy at a distance $\gtrsim \Gamma^{2}r_{\rm g}$, where $r_{\rm g}$ is the gravitational radius.
This scenario works for most of the cases of variability in optical, X-ray, and even $\gamma$-ray bands, together with those models referring to the jet spiral structure, precession, or geometric effects in the jet \citep{Gopal1992AA, Camenzind1992AA}, all the blazar variability seems mostly well explained.
Minutes time scale variability observed in the TeV band \citep{Aharonian2007ApJ, Albert2007ApJ} demands every efficient particle acceleration/dissipation mechanism \citep{Shukla2020NatCo, Wang2022PASP} or extremely small emission zone.
Although the variability mechanisms have been explored and discussed by many people, the initial seed instability of arising blazar variabilities, in different time scales, was few discussed.

A two-fluid model, which was proposed by \citep{Sol1989MNRAS}, coupled with the Kelvin–Helmholtz instability could have a chance.
In this scenario, two fluids (one fluid is a nonrelativistic jet with electron-proton plasma, and another one is a relativistic jet with an electron-positron plasma) with different speeds and compositions, the Kelvin–Helmholtz instability occurs at the junction of the two jet components and generate significant disturbances \citep{Romero1995ApSS, Cai2022ApJS}.
But, this instability only happens when the magnetic field is weaker than the critical magnetic field strength
\begin{equation}
    B_{\rm c} = [4\pi N m_{\rm e}c^{2}(\Gamma^{2}-1)]^{1/2} \Gamma^{-1},
\end{equation}
where $N$ is the particle number density that we obtained via equation \ref{n}, $\Gamma$ is assumed to be equal to $\delta$.
We calculate the $B_{\rm c}$ for FSRQs and BL Lacs in our sample, and an average value replacement is applied to those sources without available $\delta$ during the calculation, the results are listed in column (12) of Table \ref{para}.
Based on our calculation, there are 682 sources (taking 36\% sources of the FSRQs and BL Lacs) that have a magnetic field weaker than its critical one.
The result suggests that there is a significant number of \textit{Fermi} blazars could show the Kelvin–Helmholtz instability in the emission zone.
There are possible consequences of these instabilities.
The instability could be amplified due to accumulation, thus showing detectable variability and the variability time scale is dependent on the energy dissipation efficiency.
Or the instability could be generated in a manner of kink, and thus show detectable violent variability.
For instance, these kink instabilities can disrupt the jet \citep{Porth2015MNRAS452, Duran2017MNRAS469} and trigger magnetic reconnection \citep{Giannios2006A&A450, Shukla2020NatCo}.
The magnetic reconnection accelerates particles, and these particles dissipate rapidly in a very short timescale, e.g., the 3C 279 gamma-ray flare \citep{Shukla2020NatCo}.

With the assumption of two-fluid model Kelvin–Helmholtz instability, our results suggest that more than one-third of the blazars are able to show this instability.
While the result is still modest if we consider observed variabilities are initially arising from the seed instabilities because almost all the observed blazars show strong variabilities although these variabilities happen in different bands and at different epochs.
Possible reasons are the following 
(1) the SED fitting results of particle number density, that we applied to calculate $B_{\rm c}$, are biased by high states of blazars;
(2) other mechanisms of seed instability, that were not considered in this work, may involve.
However, one should keep in mind that even if all the blazars can generate instabilities, the strong and violent variability only happens occasionally because most of the instabilities fade before they are amplified to generate significant variabilities.

\subsection{The location of the emission region}
The location of the emission region is one of the essential properties in blazars.
In the framework of the one-zone leptonic model, emissions in different bands are believed to be radiated in the same region.
However, the location of the blazar emission region is controversial.
The most common method to determine the location of the emission region is by assuming a constant jet geometry and taking the full jet cross section as the emission region diameter, then obtaining the distance between the emission region and the SMBH through the basic trigonometric relations.
\citet{Foschini2011} used the $\sim$2 yr of Fermi-LAT observations to study the locations of several blazars and found the location should be within the BLR.
While the expected spectral cut-off, arising from the photon-photon pair annihilation of $\gamma$-rays with the helium Lyman recombination continuum within the BLR \citep{Poutanen2010}, at the GeV band is not always observed.
Not even the cut-off could also be explained by the other consequence, e.g., a break on the EED suggested by \citet{Dermer2015}.
Furthermore, the detection of TeV emission from blazars suggests the TeV emission region should locate outside the BLR because of the severe attenuation, the interactions between TeV photons and the photons in the BLR, would not allow us to observe the TeV emission from blazars.

In this work, we made our research on the location of the blazar emission region.
We applied a timescale of 1 day, which was suggested by \citet{Nalewajko2013MNRAS} for as a typical variability timescale in the source frame in the \textit{Fermi} $\gamma$-ray band, 
this timescale has been used in many works \citep[e.g.,][]{Ghisellini1998MNRAS301, Nalewajko2013MNRAS, Fan2013, Chen2018, Pei2022ApJ}.
One can set an upper limit on the emission region size via inequality \ref{r}, thus its distance from the SMBH can be estimated via equation \ref{Eq_psi}.
An approximation, which is assuming the viewing angle to be equal to the jet semi-opening angle, is proposed and employed in this work.
We notice that most of our sources with semi-opening angles of $\tan \! \phi <0.25$, see the upper panel of Fig. \ref{dis_d_em}.
Our result is partly consistent with the range for the semi-opening angle, which is suggested as $0.1< \tan \! \phi <0.25$ \citep{Ghisellini2009, Dermer2009}.
Meanwhile, there are 116 of 147 FSRQs (taking 78.9\%) and 20 of 35 BL Lacs (taking 57.1\%) with $\tan \! \phi \leq 0.1$.
In this case, our result suggests that maybe the viewing angle is smaller than the actual semi-opening angle and the line of sight should lie within the jet cone.

\begin{figure}
\centering
\includegraphics[scale=0.60]{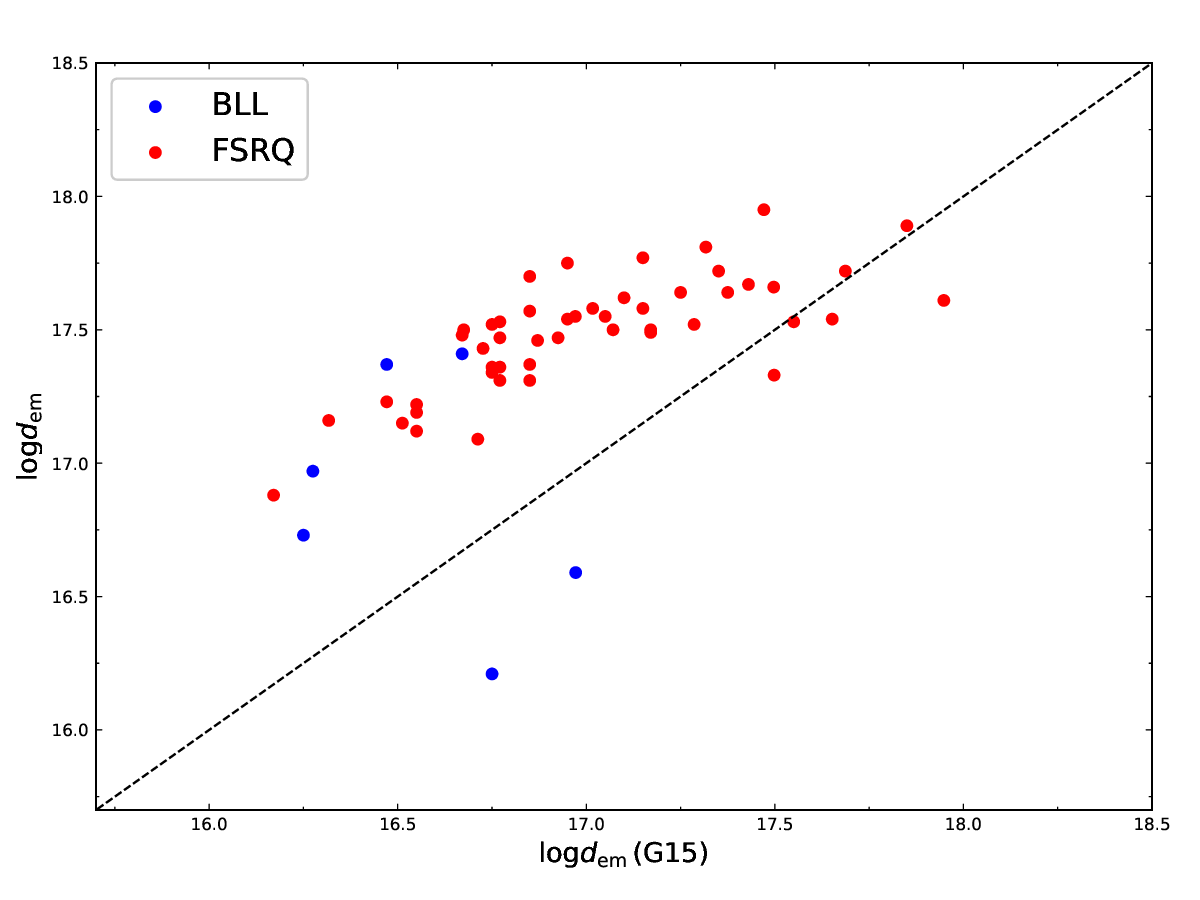}
\caption{The comparison of $\log d_{\rm em}$ from this work and from \citet{Ghisellini2015MNRAS}}
\label{dis_com}
\end{figure}

As we can see from the bottom panel of Fig. \ref{dis_d_em}, the distributions show that the emission region is located at a distance of $1.68 \times 10^{14} \, {\rm cm}$ to $6.61 \times 10^{19} \, {\rm cm}$, corresponds to $5.4 \times 10^{-5} \, {\rm pc}$ to $21.3 \, {\rm pc}$ for all blazars.
This large variance of 6 orders of magnitude directly arises from the variance of the viewing angle, from 0.04 for 4FGL J2035.4+1056 to 84.12 for 4FGL J0113.7+0225.
The mean values of $\log d_{\rm em}^{\rm B}=17.37$ (0.076 pc) for BL Lacs and $\log d_{\rm em}^{\rm F}=17.49$ (0.1 pc) for FSRQs are very close to each other, and the result of K-S test confirmed that they are from the same distribution.
Figure \ref{dis_com} shows a comparison of $\log d_{\rm em}$ from the present work and from \citet{Ghisellini2015MNRAS}, in which they estimated the $\log d_{\rm em}$ for 221 sources and gave an average value of $\sim 1\times 10^{17}\, {\rm cm}$ (0.03 pc), 
the comparison also shows that the $\log d_{\rm em}$ from the present work is larger than those from their work.
The comparison result could arise from an underestimated semi-opening angle and thus an overestimated distance from the SMBH.
In addition, different methods we employed to estimate the $\log d_{\rm em}$ could also arise the discrepancy.

Nevertheless, we check the emission region's relative location with the BLR and with the DT.
The distances from the BLR and from the DT to the SMBH are assumed to scale with the square root of the accretion disk luminosity, 
\begin{equation}
    d_{\rm BLR} = 10^{17}L_{\rm disk,\, 45}^{1/2} \, {\rm cm},
\end{equation}
and
\begin{equation}
    d_{\rm DT} = 2.5\times10^{18}L_{\rm disk,\, 45}^{1/2} \, {\rm cm},
\end{equation}
where the $L_{\rm disk,\, 45}$ is the accretion disk luminosity $L_{\rm disk}$ in units of $10^{45} \, {\rm erg \cdot s^{-1}}$ \citep{Ghisellini2008}.
The accretion disk luminosity can be estimated by the BLR luminosity $L_{\rm disk} \simeq 10 L_{\rm BLR}$ \citep{Calderone2013}.
We manage to collect the $L_{\rm BLR}$ from \citet{Xiao2022ApJ_1} for 143 of the 182 blazars and list them in Table \ref{distance}, we find that there are 66 (taking 46.2\%) sources with emission regions located within the BLR, 63 (taking 44.0\%) sources with emission regions located between the BLR and the DT, and 14 (9.8\%) sources with emission regions located beyond the DT.
Locating the emission region much further out of the DT is not appropriate because there are no important sources of external photons.
The reason for the 14 sources with overestimated $d_{\rm em}$ is that small semi-opening angles, less than 0.1, are employed.

To sum up, our results suggest about half of the blazars with the emitting regions $d_{\rm em} < d_{\rm BLR}$ and another half with the emitting regions $d_{\rm BLR} < d_{\rm em} < d_{\rm DT}$.
This is clearly inconsistent with the result suggested in \citet{Ghisellini2014}, in which they suggested 85\% sources with emission region located within the BLR and 15\% sources with emission region located between the BLR and the DT.
This discrepancy could be caused mainly by two reasons, namely, the sample size and the methods of estimating the $d_{\rm em}$.
While the method influence can not be the main reason because the deviation caused by one particular method should evenly affect $d_{\rm em}$ for all sources in the sample and should not significantly change the portion of $d_{\rm em}$ within the BLR or within the DT.
The best way to eliminate this discrepancy is to compile a larger/complete sample of blazars to estimate the distance from the SMBH to the emission region and determine the emission region's relative location to the BLR or the DT.

\section{Conclusions} \label{sec:conclusion}
In order to study the property of the blazar emission region and put constraints on the corresponding parameters, we compiled a sample of 2708 \textit{Fermi} blazars with available broadband SED features.
A fraction of the sources in our sample also has available redshift, Doppler factor, viewing angle, and BLR luminosity.
With the above-mentioned information, we calculated the magnetic field ($\log B$) and electron energy ($\log \gamma_{\rm p}$) for certain type blazars (FSRQs and BL Lacs) and provided ranges of $\log B$ and $\log \gamma_{\rm p}$ for BCUs, electron energy density ($\log U_{\rm e}$), the energy ratio ($\log (U_{\rm e}/U_{\rm B})$), and the critical magnetic strength ($B_{\rm c}$) to study the jet launching and seed variability.
A distance ($d_{\rm em}$) from the SMBH to the emission region is obtained for a subsample of 182 sources and used to discuss the emission region location compare to the BLR.

Our main conclusions are as follows:
(1) The FSRQs show lower electron energy than the BL Lacs, we suggest to divided FSRQs and BL Lacs with $\log \gamma_{\rm p}=3.20 \pm 0.01$.
Besides, FSRQs show trends of a stronger magnetic field and smaller electron-to-magnetic energy ratio than the BL Lacs;
(2) Our results suggest that the BL Lacs with a mean value of $\log (U^{\rm B}_{\rm e} / U^{\rm B}_{\rm B}) =0.98$, may fulfil the equipartition between magnetic field energy density and the electron energy density, while the FSRQs with a mean value of $\log (U^{\rm F}_{\rm e} / U^{\rm F}_{\rm B}) = -6.45$ is away from the condition $\log (U_{\rm e} / U_{\rm B}) \sim 0$; 
(3) Comparing the $B$ and $B_{\rm c}$, we find 682 blazars with $B$ samller than the $B_{\rm c}$ and suggest these sources are all candidates of showing Kelvin–Helmholtz instability.
But we note that our result is an old-line and the rest of the blazars could also have a chance to show this instability;
(4) Our result of the location of the blazar emission region is particularly novel.
The result suggests both FSRQs and BL Lacs with emission region at the distance of $\sim$0.1 pc.
And we find that about half of the blazars with emission regions within the BLR, while another half sources with the region located between the BLR and the DT.

\begin{acknowledgments}
J. H. Fan thanks the support from the NSFC (NSFC U2031201, NSFC 11733001, U2031112), Scientific and Technological Cooperation Projects (2020-2023) between the People's Republic of China and the Republic of Bulgaria, Guangdong Major Project of Basic and Applied Basic Research (Grant No. 2019B030302001), the science research grants from the China Manned Space Project with NO. CMS-CSST-2021-A06, and the support for Astrophysics Key Subjects of Guangdong Province and Guangzhou City. 
This research was partially supported by the Bulgarian National Science Fund of the Ministry of Education and Science under grants KP-06-H38/4 (2019), KP-06-KITAJ/2 (2020) and KP-06-H68/4 (2022).
H.B. Xiao acknowledges the support from the National Natural Science Foundation of China (NSFC 12203034) and from the Shanghai Science and Technology Fund (22YF1431500).
\end{acknowledgments}

%


\bibliography{lib}{}
\bibliographystyle{aasjournal}



\end{document}